\documentclass[mathleft
]{an}
\usepackage{graphicx}
\usepackage{times}
\begin{document}

\Pagespan{789}{}
\Yearpublication{2009}%
\Yearsubmission{2009}%
\Month{11}%
\Volume{999}%
\Issue{88}%

\title{Smooth, undisturbed dwarf spheroidal galaxies in
the Perseus Cluster core: Implications for dark matter content}

\author{S.J. Penny\inst{1}\fnmsep\thanks{
  \email{ppxsp@nottingham.ac.uk}\newline}
\and  C.J. Conselice\inst{1}
\and S. De Rijcke\inst{2}
\and E.V. Held\inst{3}
}
\titlerunning{Dark matter in cluster dwarfs}
\authorrunning{S.J. Penny et al.}
\institute{School of Physics $\&$ Astronomy, University of Nottingham, Nottingham NG7 2RD, UK
\and 
Sterrenkundig Observatorium, Universiteit Gent, Krijgslaan 281, S9, B-9000 Gent, Belgium
\and
Osservatorio Astronomico di Padova, INAF, Vicolo Osservatorio 5, I-35122 Padova, Italy}

\received{August 2009}
\publonline{later}


\abstract{%
Using deep HST/ACS observations of the core of the Perseus Cluster, we identify a large population
of dwarf elliptical galaxies down to M$_{V} = -12$. All these dwarfs are remarkably smooth in appearance, showing no evidence
 for internal features that could be the result of tidal processes or star formation induced by the cluster
potential. Based on these observations and the relatively large sizes of these dwarfs, we argue that at least some must have a large dark matter component to prevent their disruption by the cluster potential.
We further derive a new method to quantify the dark matter content of cluster dSphs without the use of kinematics, which are impossible to obtain at these
distances. We find that mass-to-light ratios for dwarfs in the core of the Perseus Cluster are comparable to those found for Local Group dSphs, ranging between M$_{\odot}$/L$_{\odot}$ $\approx$ 1 and 120. This is evidence that dwarf spheroidals reside in dark matter subhalos that protect them from tidal processes in the cores of dense clusters.} 

\maketitle

\section{Introduction}

Dwarf elliptical (dE) (M$_B > -18$) galaxies and the fainter dwarf spheroidal (dSph) galaxies (M$_B$ $> -14$) are the most common galaxy type in the Universe, and are furthermore believed to be among the most dark matter dominated objects. Typical mass-to-light ratios (M/L) for Local Group dSphs are $\sim$100 (e.g. Strigari et al. 2007). The role of dark matter in galaxy formation can thus potentially be obtained via observations of these low mass galaxies (Conselice, Gallagher \& Wyse 2003). 

One method to measure the dark matter content, and therefore M/L ratios, of cluster dwarfs is to determine their internal velocity dispersions using kinematic measures of individual stars. However, the large distances to galaxy clusters makes it technically impossible to resolve the stellar populations of most cluster dEs/dSphs, therefore kinematics cannot be easily used to find the dark matter content of such galaxies. 

We develop a new method for determining the dark matter content of dwarf elliptical galaxies in the dense cluster environment without the use of kinematics. This method is based on the size of the dwarf, its position in the cluster, and the smoothness and symmetry of its light profile. We measure the M/L ratios for dwarfs in the Perseus Cluster by finding the minimum total mass that each galaxy must have to prevent it from being disrupted by the cluster potential at its current projected distance from the cluster centre. We also test how much dark matter the dwarfs in our sample require to survive in the  Perseus Cluster at their likely pericentric distance.

\section{Observations and cluster membership}

As part of a general study of the dwarf galaxy population of the Perseus Cluster, we have obtained deep \textit{Hubble Space Telescope (HST)} Advanced Camera for Surveys (ACS) observations of the core of the Perseus Cluster. Perseus is one of the nearest rich galaxy clusters, with a redshift $v = 5366$ km s$^{-1}$ (Struble \& Rood 1999), and a distance $D = 72$ Mpc. These observations target five fields in the F555W and F814W bands ($V$ and $I$ respectively), with exposure times of one orbit per field. 

Our HST/ACS imaging contains 6 dwarfs with spectroscopicly confirmed cluster membership (Penny \& Conselice 2008), and we identify by eye all galaxies with similar morphologies to these confirmed cluster members. We identify 29 dwarfs in total, with a selection of these dwarfs presented in Fig. 1. All the dwarfs we identify are remarkably smooth and symmetric in appearance, without evidence for internal substructure or tidal features, suggesting they have not been tidally disrupted by the dense cluster environment.

\begin{figure}
\includegraphics[width=82mm]{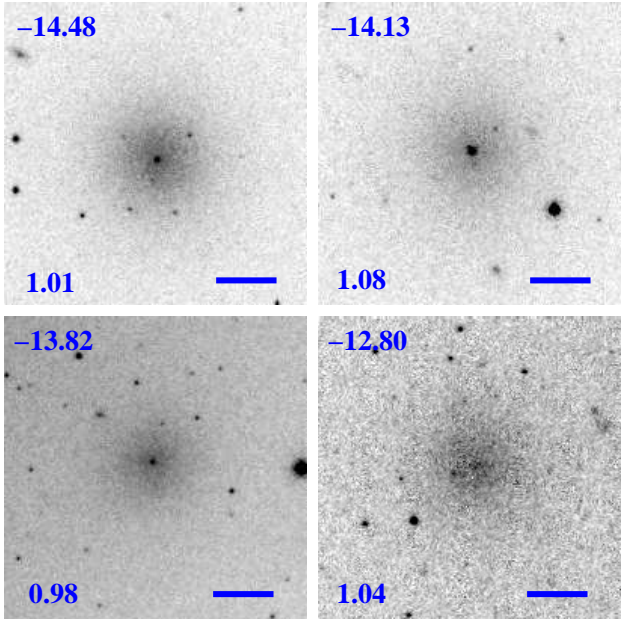}
\caption{Selection of Perseus dwarfs used in this study. The upper left number is the value of M$_{\rm{V}}$ for the dwarf, and the lower left number the ($V-I$) colour. The solid bar is 2'' in length.}
\end{figure}

\section{Determination of dark matter content}

The smoothness of our dwarf sample is quantified via parametric fitting of their two-dimensional light distributions, and through non-parametric structural parameters. Using \\our F814W band images, we extract surface brightness, position angle and ellipticity profiles for all dwarfs. All the dwarfs have low S\'{e}rsic indices ($n < 1$), showing they have near exponential light profiles, typical for dwarf ellipticals (De Rijcke et al. 2009). In general the ellipticities are constant with radius, with an average ellipticity for this sample of $<\epsilon> = 0.18 \pm 0.10$, showing these galaxies are roughly round systems. The position angles for these galaxies are also constant with radius, demonstrating that isophotal twists are not present in these dwarfs. 

To determine the non-parametric structures of our dwarf sample, we use the concentration, asymmetry and clumpiness (CAS) parameters (Conselice 2003). The concentration index is the ratio of the radii containing 80\% and 20\% of the galaxy's light, the asymmetry index measures the deviation of the galaxy's light from perfect 180$^{\circ}$ symmetry, and the clumpiness index describe how patchy the distribution of light within a galaxy is. The dwarfs in this study are found to be very smooth, with very low asymmetry and clumpiness values (A $= 0.03 \pm 0.04$, S $= 0.02 \pm 0.09$) throughout their entire structure, showing no evidence for internal features or star formation that could be the result of tidal processes in the cluster environment. 

We imply from the smoothness of the dwarfs that they likely have a large dark matter component to prevent tidal disruption by the cluster potential. However, as mentioned the large distance to the Perseus Cluster makes it difficult to obtain internal velocity dispersions for these galaxies using spectroscopy. Therefore we derive a new method to determine the dark matter content of these dwarfs by finding the minimum mass they must have to prevent their tidal disruption by the cluster potential.

To determine the dark matter content of our dwarfs, we assume they are spherical, with no substructure, and that Newtonian gravity is valid for such galaxies. Along with these assumptions, the main properties needed to measure the dark matter content of dEs are their sizes and their distances from the centre of the Perseus Cluster, which we take to be at the coordinates of NGC 1275. The radii of the dwarfs are taken to be their Petrosian radii, unless the dwarf is near a bright star or galaxy, where the dwarf is excluded, leaving a sample of 25 dwarf ellipticals. It is likely that the dark matter haloes of the dwarfs extend beyond these Petrosian radii, placing a lower limit on the total masses of these galaxies. 

In order to remain intact, the dwarfs must be sufficiently massive to withstand the tidal forces exerted on them by the cluster potential. Assuming the dwarfs are on radial orbits, the dwarf will be subject to the largest tidal forces it will experience during its orbit at its pericenter, and this will be the point at which its total mass will need to be highest to prevent its tidal disruption. No dwarfs exist within a 35 kpc radius of the cluster centred galaxy NGC 1275, so we take this to be the pericentric, and therefore minimum distance from the cluster centre at which a dwarf can survive the cluster potential without disruption. 

\subsection{Cluster mass}

To determine total masses for the dwarfs in our sample, we need to know the total mass of the cluster interior to each dwarf. Mathews, Faltenbacher and Brighenti (2006) provide a model of the acceleration due to gravity in the Perseus Cluster, based on \textit{XMM-Newton} and \textit{Chandra} X-ray observations. The acceleration due to gravity is modelled by combining a Navarro, Frenk and White (NFW) dark halo and a stellar contribution from NGC 1275 [See Mathews et al. (2006) and Penny et al. (2009) for details].

The acceleration due to gravity can then be converted to an effective mass enclosed in a radius $R$, by $M_{cl}(R)=$ $\\g(R)R^{2}/G$. The model of the cluster mass is presented in Fig. 2. Once the mass of the cluster out to the distance of the dwarf is known, the mass the dwarf required to prevent disruption by the cluster potential can be calculated. 

\subsection{Dwarf mass measurements}

The above information is used to derive tidal masses for our dwarfs. As a dwarf orbits the cluster centre, a star on the surface of the dwarf will become detached by the cluster tidal forces unless the dwarf has sufficient mass to prevent this occuring. We therefore estimate the minimum tidal mass the dwarf must have to prevent disruption by the cluster potential. We use King (1962) to derive the mass of a dwarf galaxy on a radial orbit in Perseus as

\begin{equation}
\ M_{\rm{dwarf}} > \frac{r_{d}^{3}M_{cl}(R)(3+e)}{R^3}
\end{equation}

\noindent where $M_{\rm{dwarf}}$ is the total mass of the dE, $r_{d}$ is the Petrosian radius of the dwarf, $M_{cl}$(R) is the mass of the cluster interior to the dwarf, $R$ is the peri-centric distance of the dwarf from the cluster centre, and $e$ is the eccentricity of the dwarf's orbit, which we take to be 0.5.

Total masses are calculated for galaxies of varying radii using equation 1. As we do not know the tidal radii of our dwarfs, we instead measure the masses the dwarfs would need to prevent them being disrupted at their Petrosian radii. The predicted masses of galaxies of different radii as a function of distance from the cluster centre are presented in Fig. 2.

\begin{figure}
\includegraphics[width=82mm]{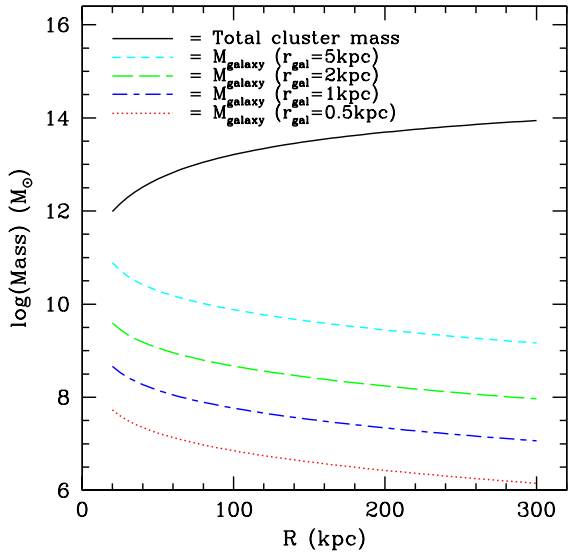}
\caption{Relationship between predicted galaxy mass and cluster centre distance for galaxies of varying radii, found using our method. The masses of modelled galaxies with radii of 0.5, 1, 2, and 5 kpc are plotted, as well as the total cluster mass, as a function of distance from the cluster centre. The model of the cluster mass is valid for the region 10 kpc $\leq R \leq$ 300 kpc.}
\end{figure}

In order to calculate stellar to total mass ratios for our dwarfs, we calculate the stellar mass for each galaxy. This is determined using the method of Bell \& de Jong (2001). The colour of each dwarf is measured within a central 1 arcsec radius, and this colour is then converted to a stellar M/L ratio by using the mass-dependent galaxy formation epoch model with a scaled-down Saltpeter IMF from Bell \& de Jong (2001). Stellar masses are then found for the dwarfs by converting their absolute magnitudes to a solar luminosity, and then multiplying by the calculated stellar M/L ratio.

\section{Results- tidal masses}

A relationship exists for Local Group dSphs such that the fainter galaxies have higher M/L ratios, with the faintest dSphs (M$_{V} = -4$) having calculated M/L ratios approaching 1000M$_{\odot}$/L$_{\odot,V}$ (Simon $\&$ Geha 2007). We investigate the relationship between M/L ratios and luminosities for dwarfs in the dense cluster environment, and measure the M/L ratios for Perseus Cluster dEs using the new method presented here. The results are shown in Fig. 3, with M/L ratios for the Milky Way dSphs from Strigari et al. (2007) included on this plot for comparison. The M/L ratios for the Perseus dwarfs are measured at both their current projected distance from the cluster centre and at a pericentric distance of 35 kpc from NGC 1275. The orbital eccentricity is assumed to be $e = 0.5$. 

\begin{figure*}[htb]
\includegraphics[width=160mm]{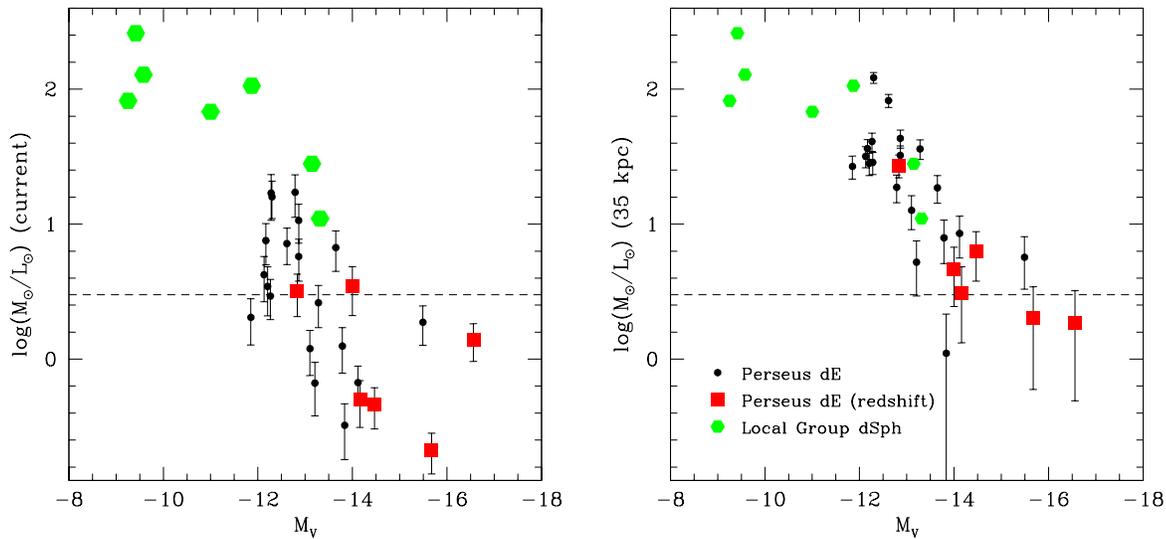}
\caption{Relationship between mass-to-light ratios and $M_{V}$ for Perseus dEs and Milky Way dSphs. The left hand plot shows M/L for the dwarfs at their current cluster position, and the right hand plot shows M/L for the dwarfs at a distance of 35 kpc from NGC 1275. Values of M/L for the Milky Way dSphs are taken from Strigari et al. (2007), and were obtained using kinematics. The dashed line in each plot is at a M/L of 3, which we take, within our errors, to be the mass-to-light ratio at which no dark matter is required to prevent the dwarf being tidally disrupted.}
\end{figure*}

Fig 3. shows that at their current positions in the cluster, 12 of the dwarfs in our sample require dark matter to prevent their disruption by the cluster potential, with the remaining dwarfs having M/L ratios smaller than 3, indicating they may not require dark matter, within our errors, at their current distances from the cluster centre. When we measure the M/L ratios that the dwarfs would require to prevent disruption at a pericentric distance of 35 kpc from the cluster centre, we find that all but three require dark matter.

\section{Discussion and conclusions}

All the dwarfs we present here are remarkably smooth, symmetrical and round in appearance, and lack internal features. The fact that these galaxies are not morphologically disturbed suggests they must contain a large proportion of dark matter to prevent tidal disruption, and we derive a new method to determine their dark matter content based on their smoothness, size and position in the cluster. 

We find a clear correlation between M/L ratios and luminosities of the dwarfs, such that the faintest dwarfs require the largest fractions of dark matter to remain bound. Like their Local Group counterparts, faint dwarf ellipticals must have a substantial proportion of dark matter to remain bound against the cluster potential, with M/L ratios ranging between 1 for the brightest dwarfs, to 120 for the faintest galaxies in our sample. This indicates that to remain intact in the cores of rich clusters, dwarf spheroidals likely reside in dark matter subhalos that protect them from the cluster potential. 

\acknowledgements
SJP acknowledges the support of a STFC studentship. SDR is a Postdoctoral Fellow of the Fund for Scientific Research- Flanders (Belgium)(F.W.O.).

\end{document}